\documentclass[longbibliography,twocolumn,prl,aps,superscriptaddress,showpacs,amsmath,amssymb,floatfix]{revtex4-1}
\usepackage{color}
\usepackage{xcolor}
\usepackage{mathrsfs}
\usepackage{amsmath}
\usepackage{graphicx}
\usepackage{dcolumn}
\usepackage{bm}
\usepackage{times}
\usepackage{amssymb}
\usepackage{float}
\usepackage{array}
\usepackage{tikz}
\usepackage{ulem}

\begin{document}

\title{Topological phase, supercritical point and emergent phenomena in extended $\mathbb{Z}_3$ parafermion chain}
\author{Shun-Yao Zhang}
\affiliation{CAS Key Laboratory of Quantum Information, University of Science and Technology of China, Hefei, 230026, People’s Republic of China}
\author{Hong-Ze Xu}
\affiliation{CAS Key Laboratory of Quantum Information, University of Science and Technology of China, Hefei, 230026, People’s Republic of China}
\author{Yue-Xin Huang}
\affiliation{CAS Key Laboratory of Quantum Information, University of Science and Technology of China, Hefei, 230026, People’s Republic of China}
\author{Guang-Can Guo}
\affiliation{CAS Key Laboratory of Quantum Information, University of Science and Technology of China, Hefei, 230026, People’s Republic of China}
\affiliation {Synergetic Innovation Center of Quantum Information and Quantum Physics, University of Science and Technology of China, Hefei, Anhui 230026, China}
\author{Zheng-Wei Zhou}
\thanks{Email: zwzhou@ustc.edu.cn}
\affiliation{CAS Key Laboratory of Quantum Information, University of Science and Technology of China, Hefei, 230026, People’s Republic of China}
\affiliation {Synergetic Innovation Center of Quantum Information and Quantum Physics, University of Science and Technology of China, Hefei, Anhui 230026, China}
\author{Ming Gong}
\thanks{Email: gongm@ustc.edu.cn}
\affiliation{CAS Key Laboratory of Quantum Information, University of Science and Technology of China, Hefei, 230026, People’s Republic of China}
\affiliation {Synergetic Innovation Center of Quantum Information and Quantum Physics, University of Science and Technology of China, Hefei, Anhui 230026, China}
\date{\today}

\begin{abstract}
Topological orders and associated topological protected excitations satisfying non-Abelian statistics have been widely explored in various platforms. 
The $\mathbb{Z}_3$ parafermions are regarded as the most natural generation of the Majorana fermions to realize these topological orders. Here we investigate 
the topological phase and emergent $\mathbb{Z}_2$ spin phases in an extended parafermion chain. This model exhibits rich variety of phases, including
not only topological ferromagnetic phase, which supports non-Abelian anyon excitation, but also spin-fluid, dimer and chiral phases from the emergent $\mathbb{Z}_2$ 
spin model. We generalize the measurement tools in $\mathbb{Z}_2$ spin models to fully characterize these phases in the extended parafermion model and map out the corresponding 
phase diagram. Surprisingly, we find that all the phase boundaries finally merge to a single supercritical point. In regarding of the rather generality of 
emergent phenomena in parafermion models, this approach opens a wide range of intriguing applications in investigating the exotic phases in other parafermion models.
\end{abstract}
\maketitle

Topological orders have been one of the major concerns in modern physics due to their potential realization of non-Abelian anyons for topological quantum computation\cite{nayak2008non, 
kitaev2003fault}. Along this line, the Majorana zero modes have been realized in experiments in semiconductor/topological insulator and superconductor hybrid 
structures\cite{mourik2012signatures, deng2012anomalous, alicea2011non, das2012zero, zhang2017quantized, he2017chiral}. This approach may be directly generalized to 
$\mathbb{Z}_k$ parafermion\cite{lindner2012fractionalizing, vaezi2013fractional,cheng2012superconducting,vaezi2017numerical,klinovaja2014time} (with $k=2$ for Majorana fermions) 
with $k$-fold ground state degeneracy, following the pioneering work by Fendley\cite{fendley2012parafermionic, fendley2014free,jermyn2014stability}. 
In these phases, the $\mathbb{Z}_3$ parafermion model is most intriguing due to its potential construction of Fibonacci anyons\cite{mong2014universal,stoudenmire2015assembling, 
vaezi2014fibonacci,alicea2016topological,vaezi2014superconducting,sreejith2016parafermion} for universal quantum computation. Recently, these parafermions are proposed to be
constructed in semiconductor and fractional quantum Hall state hybrid structures\cite{mong2014universal,alicea2016topological, vaezi2014superconducting, 
barkeshli2014synthetic, clarke2013exotic}.

In this work, we explore the emergent phenomena in the parafermion models. We consider an extended parafermion $\mathbb{Z}_3$ model, which is mapped to a $\mathbb{Z}_3$ clock model with next 
nearest neighboring (NNN) interaction. In the presence of strong Zeeman field, the clock model can be projected to a conventional $\mathbb{Z}_2$ spin model, giving rise to emergent 
spin-fluid (SF), dimer and chiral phases. This model exhibits rich variety of phases, which are characterized using various tools directly generalized from $\mathbb{Z}_2$ spin models.
We map out the whole phase diagram and find that the topological ferromagnetic parafermion (FP) phase is greatly enhanced in the presence of ferromagnetic interaction between NNN sites.
Strikingly, all the phase boundaries finally merge to a single supercritical (SC) point. This approach lays foundation for understanding exotic phases in other parafermion models.

\begin{figure}
    \centering
    \includegraphics[width=0.4\textwidth]{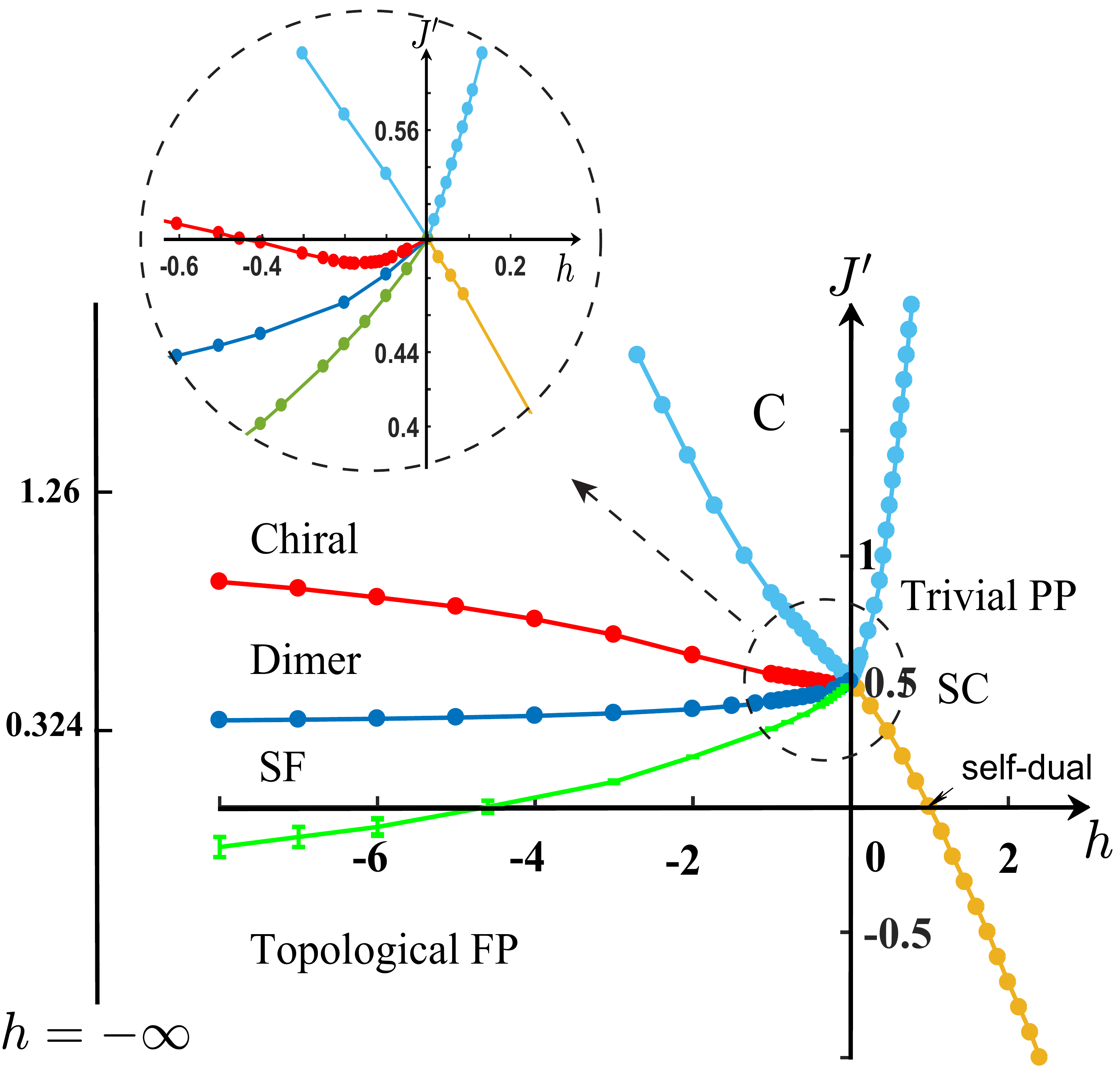}
    \caption{Phase diagram for the extended parafermion model. The abbreviations for the phases can be found in main text. All the phase boundaries finally merge to the
    supercritical (SC) point, which can be determined exactly in the classical Potts model. The left vertical axis shows the two phase boundaries 
    for the extended XX model from Refs. \onlinecite{nomura1994critical, furukawa2010chiral},  corresponding to the limit $h\rightarrow -\infty$ in our model.}
    \label{fig-fig1}
\end{figure}

{\it Model and Hamiltonian}. We consider the following extended $\mathbb{Z}_3$ parafermion chains, 
\begin{eqnarray}
    H = \mathcal{H}_0 + \mathcal{V},
    \label{eq-gparafermion}
\end{eqnarray}
where $\mathcal{H}_0 = -\omega^2 (J\sum_{j=1}^{L}\alpha_{2j}^\dagger \alpha_{2j+1} + h\alpha_{2j-1}^\dagger \alpha_{2j}) + \text{h.c.}$, $\mathcal{V} = 
J' \sum_{j} \omega \alpha_{2j}^\dagger \alpha_{2j+1} \alpha_{2j+2}^\dagger \alpha_{2j+3} + \text{h.c.}$, with $\omega = e^{i2\pi/3}$ and $\alpha_j$ are parafermions satisfying $\alpha_j^3 = 1$, 
$\alpha_j^\dagger = \alpha_j^2$ and $\alpha_i \alpha_j = \alpha_j\alpha_i \omega^{\text{sgn}(i-j)}$, $J$ and $h$ correspond to the pairings in even-odd and odd-even sites, respectively. 
In this work, we focus on region with $J > 0$\cite{fendley2012parafermionic,clarke2013exotic,zhuang2015phase}. This Hamiltonian can be mapped to the following $\mathbb{Z}_3$ clock model 
through Jordan-Wigner transformation 
$\alpha_{2j -1} = \prod_{k \le j-1} \tau_k \sigma_j$ and $\alpha_{2j} = \omega \sigma_j \prod_{k \le j} \tau_k$, which yields,
\begin{equation}
    H = -J \sum_{i=1}^L \sigma_i^\dagger \sigma_{i+1} + J' \sum_{i} \sigma_i^\dagger \sigma_{i+2} - h \sum_i \tau_i + \text{h.c.}.
    \label{eq-clock}
\end{equation}
We see that the odd-even pairing contributes to an effective Zeeman field, and the four-body 
interaction contributes to the NNN spin-spin interaction. Here the $\sigma$ and $\tau$ operators at the same site satisfy, 
$\sigma_i^3 =\tau_i^3 =1$, $\sigma_i\tau_i = \omega \sigma_i\tau_i$, $\sigma_i^\dagger = \sigma_i^2$ and $\tau_i^\dagger = \tau_i^2$,
and all operators commute, similar to those in the $\mathbb{Z}_2$ spin models, between different sites; see supplementary material
for more details\cite{supp}. In regarding of the rich phases in extended $\mathbb{Z}_2$ spin models, we expect Eq. \ref{eq-clock} to harbor rich phases besides the widely studied 
topological phase. Hereafter, $h$ is termed as Zeeman field for its similar role to magnetic field.

{\it General physics in some special points}. Before discussing the numerical phase diagram in details in Fig. \ref{fig-fig1}, we first focus on the basic physics in several 
interesting limits. When $J' = 0$,  the system can be invariant up on a self-dual transformation, $\mu_{j}=\prod_{k \le j}\tau_{k}$ and $\nu_{j}=\sigma_{j}^{\dagger}\sigma_{j+1}$,
via which Eq. \ref{eq-clock} becomes,
\begin{equation}
    H = -h \sum_{i} \nu_{i}^\dagger -J \sum_{i} \mu_{i}^\dagger \mu_{i} + \text{h.c.}.
\end{equation}
Here, $\mu_i$ and $\nu_i$ have the same algebra as $\sigma$ and $\tau$\cite{supp}. This transformation means that the model is invariant when $h = J$, giving rise to a self-dual 
critical point. From Eq. \ref{eq-gparafermion}, we see that when $|h| \ll J$, the even-odd pairing channel is dominated, leaving $\alpha_1$ and $\alpha_{2\text{L}}$ unpaired. This case 
corresponds to the physics in topological phase regime. On the other hand when $h \gg J$, the odd-even pairing channel is dominated, yielding a trivial phase. 
Therefore the self-dual point defines boundary between a trivial phase and a topological phase. This critical point was studied in literatures\cite{fendley2012parafermionic,
zhuang2015phase,li2015criticality}. 

When $h = 0$, Eq. \ref{eq-clock} is reduced to the three-state Potts model. Let us define $\sigma | s\rangle = \omega_s |s\rangle$ for $s = \{\uparrow, \searrow, \swarrow\}$, with corresponding 
eigenvalues are $\omega_s = \{1, \omega, \omega^2\}$, respectively. When $J'\sim 0$, the ground states are threefold degenerate with 
corresponding ferromagnetic wave functions,
\begin{equation}
    |g_1\rangle = |\uparrow\rangle^{\otimes L}, \quad 
    |g_2\rangle = |\searrow\rangle^{\otimes L}, \quad 
    |g_3\rangle = |\swarrow\rangle^{\otimes L}. 
    \label{eq-directspinproduct}
\end{equation}
For this reason, this phase is defined as topological FP phase. The corresponding ground state energy is $E_g^{1} = -2LJ + 2LJ'$. On the other hand, 
when $J' \gg J$, the ground state wave function 
may be written as $|s_1s_1s_2s_2 s_3s_3 \cdots \rangle$, where $s_i \in \{\uparrow,\searrow,\swarrow\}$ and $s_i \ne s_{i+1}$, with $E_g^2 = -LJ/2 -LJ'$. 
In this case, the ground state is $\mathcal{N} = 3\times 2^{L/2-1}$-fold degenerate, which approaches infinite in thermodynamic limit. The crossover between these two phases is determined 
by $E_g^1 = E_g^2$, and yields $J' = J/2$. The similar critical point can be found in the classical Ising model with NNN interaction\cite{IsingC}, in which both phases 
are two-fold degenerate. We will show that this infinite accidental degeneracy will turn to critical in the presence of Zeeman field $h$.

\begin{figure}
    \centering
    \includegraphics[width=0.45\textwidth]{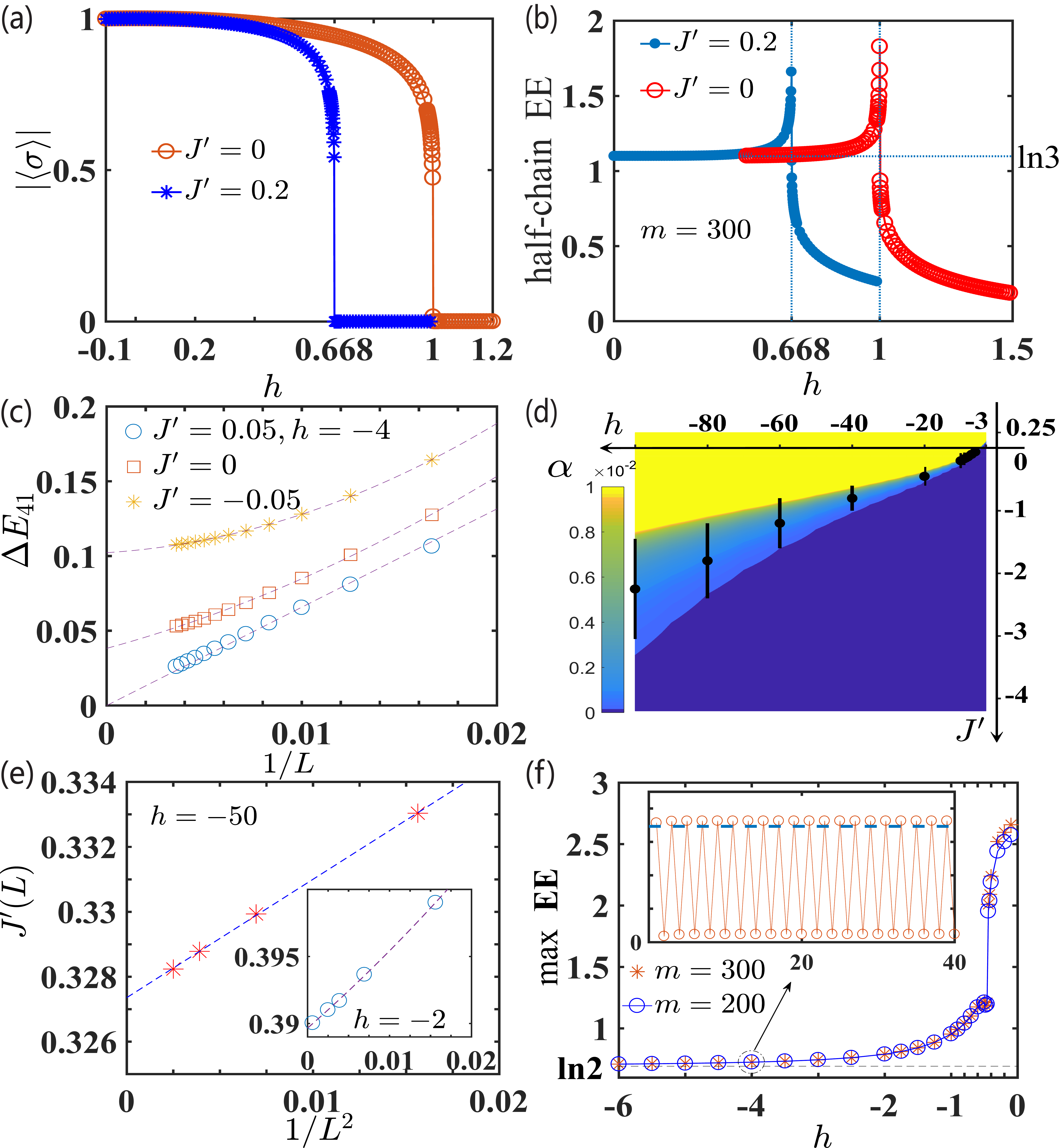}
    \caption{(a) and (b) show the ferromagnetic order and EE across the boundary between FP and PP phases. (c) Scaling of excitation gap $\delta E_{41} = E_4 - E_1$ as a function of chain length when across the 
    boundary between FP phase and SF phase. (d) Fitted scaling exponent $\alpha$ in SF and FP phases. Symbols are phase boundaries determined by vanishing of excitation gap $\delta E_{41}$. 
    (e) Phase boundary between SF phase and dimer phase by energy level crossing in the first excited bands. (f) Maximal EE in the central of chain as a function of $h$. Inset shows a typical EE 
    in the dimer phase regime. }
    \label{fig-fig2}
\end{figure}

The key insight is that the system is not always threefold degenerate when $h < 0$. When $h \rightarrow  -\infty$, the ground state in each site 
should be twofold degenerate in eigenvectors of $\tau+\tau^\dagger$\cite{supp}. In this case, the system should occupy only the lowest two states, and we obtain the following 
emergent $\mathbb{Z}_2$ spin model,
\begin{equation}
    H = -J \sum_{i} s_i^\dagger s_{i+1} + J' \sum_{i} s_i^\dagger s_{i+2} + \text{h.c.}, 
    \label{eq-XXmodel}
\end{equation}
where $s_i = {1\over 2}(s_i^x - i s_i^y)$ and $s_i^\dagger = {1\over 2}(s_i^x + i s_i^y)$, with $s_i^{x,y}$ denoting the spin-${1\over 2}$ Pauli matrices. 
This model hosts a number of interesting phases, as unveiled in Refs. \cite{nomura1994critical,nomura1993phase,totsuka1998magnetization,jafari2007phase}. Especially, 
it is relevant to two well-known results. When $J' = J/2$, it corresponds to the Majumdar-Ghosh (MG) dimer model\cite{majumdar1970antiferromagnetic, majumdar1969next}, 
in which the ground states are described by product of singlet dimer. 
Notice that in the original MG model isotropic antiferromagnetic Heisenberg model was considered, while in Eq. \ref{eq-XXmodel}, it only has two components. Following Ref. \cite{kumar2002quantum}, we find that when $J > 0$, the ground states are still described by exact singlet dimers with twofold degeneracy\cite{supp}.
Moreover, when $J' = 0$, it corresponds to the nearest neighbor XX model, which can be reduced to the single particle fermion model after a Jordan-Wigner transformation. This
is a gapless phase with central charge $c = 1$\cite{eisert2010colloquium}. These two phases are totally different from the topological FP phase 
with threefold degeneracy, thus there should be another phase boundary in regime $h < 0$. These results provide important glimpse to the novel phases in our model.

{\it Topological phase and topological phase transition}. Our phase diagram in Fig. \ref{fig-fig1} is determined by exact diagonalization (ED) and density matrix renormalization group (DMRG) methods. We present our results in unit of $J$ (set $J =1$). Firstly and most importantly, we focus on the properties of topological FP phase and characterize it using several different approaches. For a finite chain, the two ends may support localized edge modes and the coupling between them breaks the three-fold degeneracy of the ground states. We have verified these features using a finite chain by ED and DMRG calculations; see \cite{supp}. The transition from FP phase to trivial phase with $J' = 0$ has been explored in previous literatures\cite{fendley2012parafermionic,
zhuang2015phase,li2015criticality}. The important tool we have introduced in this work to understand this phase transition is based on the so-called ferromagnetic order, $\Delta = |\langle \sigma \rangle|$. For the spontaneous symmetry breaking phase in Eq. \ref{eq-directspinproduct}, $\Delta = |\langle s| \sigma |s \rangle| = 1$. Nevertheless, in the trivial phase with $h \gg J$, when only the lowest ground state $|0\rangle = {1 \over \sqrt{3}}(|\uparrow\rangle + |\searrow\rangle + |\swarrow\rangle)$ is occupied\cite{supp}, we have $\Delta = |\langle 0|\sigma|0\rangle| = 0$. This phase is resemblance to the paramagnetic phase in the transverse Ising model, thus it is termed as paramagnetic parafermion (PP) phase. In Fig. \ref{fig-fig2}a, we show that this order suddenly drops to zero at the phase boundary. With this method, we can precisely determine the boundary between FP phase and PP phase in Fig. \ref{fig-fig1}.

\begin{figure}
    \centering
    \includegraphics[width=0.45\textwidth]{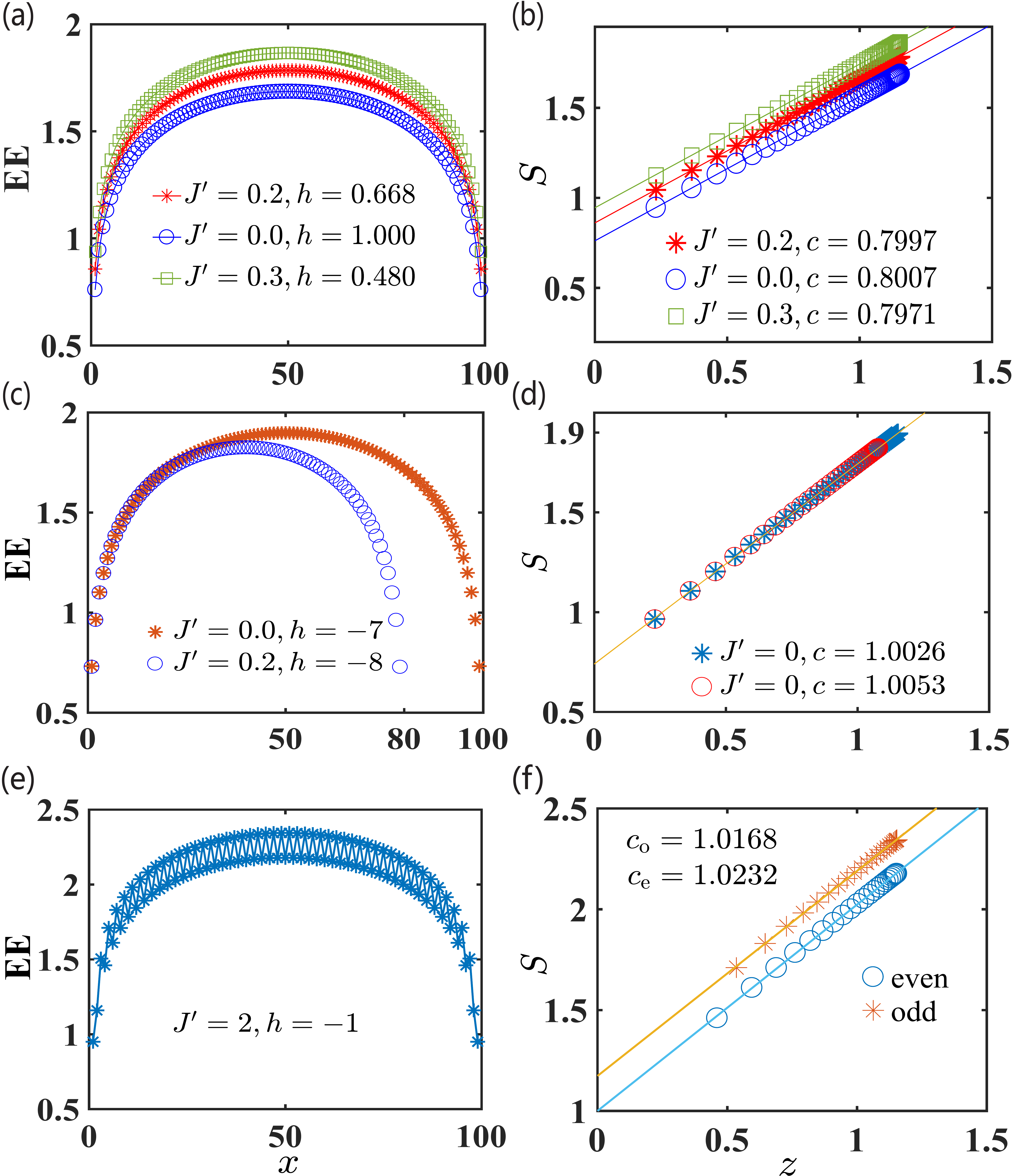}
    \caption{(a) and (b) show EE and central charge $c$ at the phase boundary between topological FP
    phase and trivial PP phase, which is critical with $c={4\over 5}$. (c) and (d) show results for SF phase with $c=1.0$. (e) and (f) show results in C phase, with
    $c_\text{o} = c_\text{e} = 1.0$ for odd chain and even chain, respectively.  In right column, $z = {1 \over 3} \ln ({L \over \pi} \sin {\pi x \over L})$. The EE is obtained 
    from DMRG with periodic boundary condition.}.
    \label{fig-fig3}
\end{figure}

We further characterize this phase transition using entanglement entropy (EE). In general, EE in a finite size with periodic boundary condition is written as\cite{calabrese2009entanglement, 
vidal2003entanglement, li2015criticality}
\begin{equation}
    S(x) = {c \over 3} \ln ({L \over \pi} \sin {\pi x \over L}) + s_0, \quad
    \label{eq-EE}
\end{equation}
where the central charge $c = 0$ in the fully gapped phase, and $s_0$ is a constant. A prefactor ${1\over 2}$ should be multiplied to the above expression 
in an open chain. The result is presented in Fig. \ref{fig-fig2}b, in which $s_0 = \ln 3$ in the FP phase regime reflects three-fold degeneracy of the ground states. In Fig. \ref{fig-fig3}a-b, we plot the 
EE (Eq. \ref{eq-EE}) as a function of $x$ at the phase boundary, indicating of criticality with $c = {4 \over 5}$. This critical phase is useful to construct the Fibonacci anyons by raising the system 
from one dimension to two dimensions for universal topological quantum computation\cite{mong2014universal,stoudenmire2015assembling,
vaezi2014fibonacci,alicea2016topological,vaezi2014superconducting,sreejith2016parafermion}. Different from the proposals in Ref. \cite{mong2014universal}, in our model a new dimension is introduced, 
providing more knobs in experiments to realize these anyons in future. Empirically, this phase boundary can be fitted well using $J' = 0.5 - 0.406h -0.094h^2$ for $h > 0$.

Next, we turn to the phase transition from the topological FP phase to the SF phase, which is gapless and critical\cite{vidal2003entanglement, peschel2004entanglement}, thus can no longer be captured by the ground state degeneracy and sharp peak in EE. This transition is marked by a transition from a fully gapped phase to a gapless phase, described by $\delta E_{n1}(L) = E_n - E_1$ for $n \ge 4$, which scales as 
$\delta E_{41} \sim L^{-1}$ (see Fig. \ref{fig-fig2}c). This criteria is used to determine the boundary between SF and FP phases, in which when $J' = 0$, $h_c \simeq -4.6$. 

The gapped phase and gapless phase have totally different behaviors from their long-range spin-spin correlations, which throughout this work is defined as $C(x) = \langle \sigma_i^\dagger \sigma_{i+x}\rangle$. In the gapless phase regime, it decays in power law as\cite{haldane1982spontaneous}
\begin{equation}
\lim_{x\rightarrow \infty} C(x) \sim  |x|^{-\alpha}.
\label{eq-sf}
\end{equation}
In the XX spin model and single particle models, $\alpha = 1/2$. This expression also holds in the gapped FP phase with long-range order, in which $\alpha = 0$.
The computed results by scaning of $J'$ and $h$ are presented in Fig. \ref{fig-fig2}d, which verify these expected features. Especially, the boundary 
determined in this way agrees well with that based on closing of gap in regime $h \in (-15, 0)$. Nevertheless, it becomes poorer and poorer with increasing of Zeeman field, 
albeit their trends are similar, due to difficulty in separating vanishing of excitation gap and states with small energy gaps. In Fig. \ref{fig-fig3}b-d, we present the EE in the SF phase regime, which yields $c = 1$, demonstrating its single particle fermionic feature even with parafermions. With these methods, we can completely determine the two phase boundaries for the topological FP phase, which merge to the supercritical (SC) point at $h=0$ and $J' = 1/2$. These results show that although strong Zeeman field regimes can destroy the topological FP phases, this phase regime can be greatly increased with ferromagnetic interaction between the NNN sites ($J' < 0$). With anti-ferromagnetic interaction ($J' > 0$), however, the long-range order is destroyed, which can be attributed to frustration mechanism.

\begin{figure}
    \centering
    \includegraphics[width=0.45\textwidth]{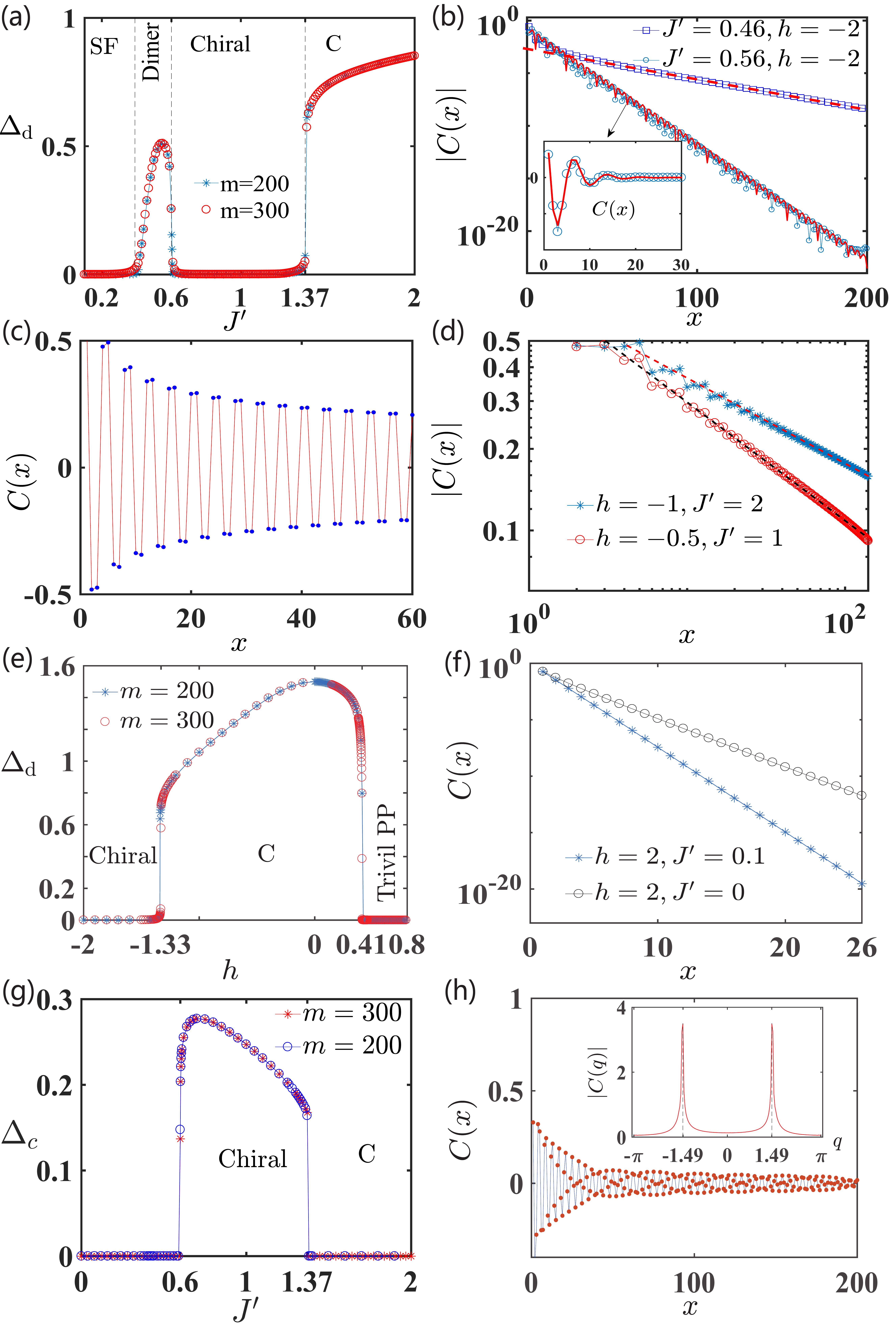}
    \caption{(a) Characterization of dimer phase and C phase using the dimer order $\Delta_\text{d}$ at $h=-2.0$ . (b) and (c) are long-range correlations in  
    dimer phase ($J^{\prime}=0.46,0.56$ for $h=-2$) and C phase ($J^{\prime}=2,h=-1$), respectively. In (b) the correlation function decays exponentially with fitted parameters $\xi = 3.97,Q=0.85$ for $h = 0.56$ and $\xi = 14.92$ for $h=0.46$. (d) In C phase, the absolute value of correlation function follows a power law decay described by Eq. \ref{eq-C} with $\alpha=0.28 (0.21)$ for $J^{\prime}=1.0 (2.0)$. 
    (e) Dimer order at $J^{\prime}= 1.0$. In (f), the long-range correlation decays exponentially with $\xi=0.578,J^{\prime}=0.1$ and $\xi=1.001,J^{\prime}=0$. (g) Chiral phase characterized by chiral order $\Delta_\text{c}$ at $h=-2.0$, which exactly equals to zero in the dimer phase and C phase. (h) Long-range correlation function in chiral phase with $J^{\prime}=2$ and $h=-4$, which is well-described by Eq. \ref{eq-chiral} with parameters: $Q=1.49,\alpha=0.69,q_0=0.067$. The inset shows the Fourier transformation of $C(x)$, and the two peaks are determined by $\pm Q$.} 
    \label{fig-fig4}
\end{figure}

{\it Properties of the emergent phases}. We next move to the emergent phases and their characterizations. In Refs. \cite{nomura1994critical,furukawa2010chiral}, a special technique was developed to investigate the phase diagram of the extended XXZ model, which hosts SF, dimer and chiral phases. We inherit these three notations in assigning the parafermion phases in Fig. \ref{fig-fig1}. When $h \rightarrow -\infty$, our model is reduced to an extended XX model (see Fig. \ref{fig-fig1}), in which the critical points between dimer and SF (chiral) phase are $J'_{\text{sf-dimer}} = 0.324$ and $J'_{\text{dimer-chiral}} = 1.26$ (see the left vertical axis in Fig. \ref{fig-fig1}). Moreover, the transition between SF and dimer phases is Kosterlitz-Thouless type and is determined by the energy level crossing between the first and second 
excited states\cite{nomura1994critical}
\begin{equation}
    \delta E_{23}(J', L) = E_3(J', L)  - E_2(J', L) = 0. 
    \label{eq-DeltaE23}
\end{equation}
With this criteria, we can find scaling of the critical point $J'(L) = J'(\infty) + A/L^2$. Two typical results based on this scaling law are presented in Fig. \ref{fig-fig2}e. When $h$ is sufficiently large, it agrees with the result in Ref. \cite{nomura1994critical}. Notice that a much longer lattice with periodic boundary condition should be used when $h$ is small, which can be implemented using DMRG method. This method yields 
the phase boundary between SF phase and dimer phase in Fig. \ref{fig-fig1}, which also merges to the SC point.

For the smooth connection between $\mathbb{Z}_3$ clock model and $\mathbb{Z}_2$ spin model, we naturally
expect the characterization of these phases by order parameters, generalized directly from the $\mathbb{Z}_2$ spin models. To this end, we define the dimer order  $\Delta_\text{d}$ and chiral order $\Delta_\text{c}$ as,
\begin{equation}
    \Delta_\text{d} = \langle \sigma_i^\dagger \sigma_{i+1} - \sigma_{i+1}^\dagger \sigma_{i+2} \rangle, 
    \Delta_\text{c} = -{i \over 2} \langle \sigma_i^\dagger \sigma_{i+1}\rangle + \text{h.c.},
    \label{eq-DD}
\end{equation}
which account for two different ways for spontaneous symmetry breaking. Specifically, the dimer order reflects the translation symmetry breaking and
the chiral order reflects the chiral symmetry breaking, {\it i.e.}, $(\boldsymbol{\eta}_i\times \boldsymbol{\eta}_{i+1})\cdot \hat{z} \ne 0$ for $\boldsymbol{\eta}_i = (\sigma_i, \sigma_i^\dagger, 0)$. 
Moreover, these phases should also exhibit different behaviors in long-range spin correlations. 

We use these tools to fully characterize all the other phases in Fig. \ref{fig-fig1}. We find that the dimer order $\Delta_\text{d}$ is nonzero in the dimer phase and C phase (Fig. \ref{fig-fig4}a), and $\Delta_\text{d} = 0$ in all other phases. Although characterized by the same order, however, these two phases have totally different behaviors in  long range correlations. In dimer phase, the correlation decays 
exponentially as
\begin{equation}
\lim_{x\rightarrow \infty} C(x) \sim \cos(Qx+q_0) \exp(-|x|/\xi),
\label{eq-exponentialdecay}
\end{equation}
where $Q$ is not necessary to be commensurate with the lattice period. This is different from that in the C phase, which decays in a power law as 
\begin{equation}
    \lim_{x\rightarrow \infty} C(x) \sim \cos(\pi x/2 - \pi/4) |x|^{-\alpha}.
    \label{eq-C}
\end{equation}
See Fig. \ref{fig-fig4}b,d for the fitted values for the parameters in these phases. 
Due to the commensuration between $C(x)$ and lattice period, this phase is termed as commensurate (C) phase. The power law decaying of $C(x)$ also indicate 
criticality breaking from infinite-fold degeneracy by finite Zeeman field. In Fig. \ref{fig-fig3}e-f, we indeed find that the central charge fitted with even chain and odd chain respectively yields 
$c_\text{o} = c_\text{e} = 1$, while the oscillating of EE reflects the nature of translation symmetry breaking. 
With this approach, we are able to precisely determine the phase boundaries between the chiral phase, dimer phase and C phase, and these two boundaries also finally merge to the SC point. 
We stress that the boundary between dimer and SF phases determined in this way is consistent with that obtained from level crossing $\Delta E_{23} = 0$; see Ref. \cite{supp}.

The dimer order can also be used to determine the phase boundary for the C phase (Fig. \ref{fig-fig4}a,e), since $\Delta_\text{d} = 0$ for the trivial PP phase with wave function $|0\rangle^{\otimes L}$. Especially, in the PP phase, the correlation decays exponentially to zero without oscillation ($Q = 0$). To gain a much deeper understanding of properties of the chiral phase, we then compute the chiral order $\Delta_\text{c}$, which is nonzero only in the regime with $\Delta_\text{d} = 0$ (see Fig. \ref{fig-fig4}a and e). In the chiral phase regime, we find that the long-range correlation decays to zero as 
\begin{equation}
\lim_{x\rightarrow \infty} C(x) \sim |x|^{-\alpha} \cos(Qx + q_0),
\label{eq-chiral}
\end{equation}
with features in combination of the incommensurate phase and gapless phase. Special concern should be payed to the boundary between the chiral and dimer phases, which will first intersect with the line $J'=1/2$ near $h = -0.44$, and then bend back to the SC point. This behavior accounts for the jump of maximum EE observed in Fig. \ref{fig-fig2}f.

{\it Conclusion and discussion}. To conclude, we investigate the topological phase and emergent fermions in the extended parafermion chain. This model exhibits rich variety of phases, including the topological FP phase, SF phase, dimer phase, chiral phase, C phase and trivial PP phase. We characterize all these phases using various methods, in which some of them are directly generalized from $\mathbb{Z}_2$ spin models. Surprisingly, we find that all the phase boundaries finally merge to a SC point. In regarding of the rather generality of emergent phenomena in parafermion models, we expect our approach as well as the measurement tools provide a general paradigm to investigate the intriguing phases in these models and unveil their intimate relations. Finally, we remark that new types of phases, such as the Haldane phase with long-range order and four-fold degeneracy, can also be realized with these $\mathbb{Z}_3$ or $\mathbb{Z}_k$ parafermions, which will be discussed elsewhere.

\bibliography{ref}

\begin{thebibliography}{41}
\bibitem [{Isi()}]{IsingC}%
  \BibitemOpen
  \href@noop {} {}\bibinfo {note} {In the classical Ising model $H = \sum_i
  -\sigma_i \sigma_{i+1} + J \sigma_i \sigma_{i+2}$, where $\sigma_i = \pm 1$,
  the phase transition point takes place at $J = 1/2$. When $J < 1/2$, the
  ground state is $|\sigma\rangle^{\otimes L}$, while for $J > 1/2$, the ground
  state is $|\sigma\sigma \bar{\sigma}\bar{\sigma}\rangle^{\otimes L/4}$ for
  $\sigma = \pm 1$ and $\bar{\sigma} = -\sigma$. In this model both phases are
  two-fold degenerate.}\BibitemShut {Stop}%
\end{thebibliography}

\end{document}